\documentclass[twocolumn,aps,prd,preprintnumbers,nofootinbib]{revtex4}
\usepackage[dvips]{graphicx}
\usepackage{bm,latexsym,amsmath,amssymb,amsfonts}

\usepackage[normalem]{ulem}

\begin{document}

\title{\uline{}Strictly static spacetimes and positive mass in the Einstein-Gauss-Bonnet theory}

\author{Tetsuya Shiromizu and Seiju Ohashi}
\affiliation{Department of Physics, Kyoto University, Kyoto 606-8502, Japan}

\begin{abstract}
In the Einstein gravity, it is well-known that strictly stationary and vacuum regular 
spacetime should be the Minkowski spacetime. In the Einstein-Gauss-Bonnet theory, we shall 
show the similar statement, that is, strictly static(no event horizon), vacuum and asymptotically flat 
spacetimes with conformally static slices are the Minkowski spacetime when the curvature corrections are 
small. 
\end{abstract}
\maketitle

\section{Introduction}

String theory predicts the higher curvature corrections to the gravitational theory at high energy \cite{HD}. 
Although we do not know the definite expression for the corrections, we know that the Gauss-Bonnet correction term 
appears in the heterotic string theory \cite{HD-GB, Metsaev}. Then one often discusses cosmology and black hole physics in the gravitational theory with the 
Gauss-Bonnet term \cite{Review-GB}, that is, the Einstein-Gauss-Bonnet theory. One striking thing is that almost 
of all fundamental theorems for the Einstein theory cannot be extended to the Einstein-Gauss-Bonnet theory. 
The singularity theorem \cite{HE}, positive mass theorem \cite{Schoen81,Witten81} and so on do not hold in general. 
About black holes, we do not know if the uniqueness theorem holds \cite{UT}. At first glance, the way to prove 
the static black hole uniqueness in the Einstein theory \cite{StaticUT} cannot be simply extended to the 
Einstein-Gauss-Bonnet case. 

There is also a simple question on the strictly stationary vacuum spacetimes. By strictly stationary spacetime, 
we mean that the timelike Killing vector exists in whole spacetime, that is, there are no event horizons. 
In the Einstein theory, it is well-known that the strictly stationary vacuum spacetimes are only the Minkowski 
spacetime \cite{Lich} (See Ref. \cite{SOS} for higher dimensions). This statement tells us that 
there is no non-trivial solution. One might have the same question in the Einstein-Gauss-Bonnet 
theory. In this paper we shall address this issue. At the same time we can briefly discuss the positive mass 
theorem because we will use it in our argument. Of course, there is no reason why the same feature holds in 
the Einstein-Gauss-Bonnet theory because the large curvature corrections may drastically 
change the structure of spacetimes. However, one can expect that they hold when curvature 
corrections are small. Indeed, we can show that the strictly static vacuum and asymptotically flat spacetimes with 
the conformally flat slices are the Minkowski spacetime in the Einstein-Gauss-Bonnet theory when the curvature 
correction to the Einstein theory is small and the Gauss-Bonnet coupling constant is positive. We note that our 
argument in this paper is relied on the perturbative treatment. 

The remaining part of this paper is organised as follows. In Sec. II, we describe the Einstein-Gauss-Bonnet 
theory and the basic equations in the static spacetimes. In Sec. III, we show that the strictly static 
spacetimes with the conformally flat and static slices are the Minkowski spacetime for certain case. Finally 
we give the summary and discussion in Sec. IV. 

\section{Static spacetimes in Einstein-Gauss-Bonnet theory}

In this section, we introduce the Einstein-Gauss-Bonnet theory and write down the field equations for the 
static spacetimes. 

We consider the gravitational theory with the Gauss-Bonnet term in $(n+1)$-dimensional spacetimes described by 
the action
%
\begin{eqnarray}
S & = & \int d^{n+1}x{\sqrt {-g}}\Bigl(R \nonumber \\
  & & ~~~~+\alpha (R_{abcd}R^{abcd}-4R_{ab}R^{ab}+R^2) \Bigr),
\end{eqnarray}
%
where $\alpha$ is the coupling constant. In the four dimensional spacetimes, the Gauss-Bonnet term does not 
affect the field equations. 

The gravitational field equations are 
%
\begin{eqnarray}
R_{ab}=-\alpha \tilde H_{ab},
\end{eqnarray}
%
where 
%
\begin{eqnarray}
\tilde H_{ab} & = & 2R_{acde}R_b^{~cde}-4R_{acbd}R^{cd}-4R_{a}^{~c}R_{bc}+2RR_{ab} \nonumber \\
 & & ~~~-\frac{1}{n-1}g_{ab}
{\cal L}_{\rm GB}
\end{eqnarray}
%
and
%
\begin{eqnarray}
{\cal L}_{\rm GB}=R_{abcd}R^{abcd}-4R_{ab}R^{ab}+R^2. 
\end{eqnarray}
%
Note that, for example, the null convergence condition $R_{ab}k^ak^b \geq 0$ 
does not hold for the tangent vector of null geodesics $k^a$ in general situations of 
the Einstein-Gauss-Bonnet theory. 
This condition is required to show some fundamental theorems like the singularity theorem, 
black hole area theorem and so on \cite{HE}. 

Here we consider the strictly static spacetimes with the hypersurface orthogonal 
timelike Killing vector, $\partial_t$. The metric of the static spacetime can be written as 
%
\begin{eqnarray}
ds^2=-V^2(x^i)dt^2+g_{ij}(x^k)dx^i dx^k, 
\end{eqnarray}
%
where $g_{ij}$ is the metric of $n$-dimensional $t=$constant space and indices $i,j$ run over 
$1, 2, \cdots, n-1$. The strictly static spacetime means that there are no Killing horizons, 
that is, $V$ is positive. In static spacetimes, the non-trivial components of the Riemann tensor 
are 
%
\begin{eqnarray}
R_{ijkl}={}^{(n)}R_{ijkl} 
\end{eqnarray}
%
and
%
\begin{eqnarray}
R_{0i0j}=VD_i D_j V,
\end{eqnarray}
%
and then those of the Ricci tensor are  
%
\begin{eqnarray}
R_{00}=VD^2V 
\end{eqnarray}
%
and
%
\begin{eqnarray}
R_{ij}={}^{(n)}R_{ij}-\frac{1}{V}D_i D_j V, \label{rij}
\end{eqnarray}
%
where ${}^{(n)}R_{ijkl}$, ${}^{(n)}R_{ij}$ and 
$D_i$ are the $n$-dimensional Riemann tensor, Ricci tensor and the covariant derivative with 
respect to $g_{ij}$ respectively. 

Now ${\cal L}_{\rm GB}$ is written as 
%
\begin{eqnarray}
{\cal L}_{\rm GB}={}^{(n)}{\cal L}_{\rm GB}+\frac{8}{V}{}^{(n)}G_{ij}D^i D^j V,
\end{eqnarray}
%
where ${}^{(n)}G_{ij}={}^{(n)}R_{ij}-(1/2)g_{ij}{}^{(n)}R$ and 
${}^{(n)}{\cal L}_{\rm GB}$ is 
%
\begin{eqnarray}
{}^{(n)}{\cal L}_{\rm GB}& := & {}^{(n)}R_{ijkl}{}^{(n)}R^{ijkl}-4{}^{(n)}R_{ij}{}^{(n)}R^{ij}+{}^{(n)}R^2 
\nonumber \\
& = & {}^{(n)}C_{ijkl}{}^{(n)}C^{ijkl} \nonumber \\
& & -4\frac{n-3}{n-2}\Bigl( {}^{(n)}R_{ij}^2
-\frac{n}{4(n-1)}{}^{(n)}R^2 \Bigr).
\end{eqnarray}
%
In the above, ${}^{(n)}C_{ijkl}$ is the $n$-dimensional Weyl tensor with respect to $g_{ij}$. 

In static spacetimes, the Hamiltonian constraint is
%
\begin{eqnarray}
{}^{(n)}R+\alpha {}^{(n)}{\cal L}_{\rm GB}=0. \label{Hami}
\end{eqnarray}
%
(See Ref. \cite{Torii} for the Arnowitt, Deser and Misner(ADM) formalism of the 
Einstein-Gauss-Bonnet theory.).

\section{Strictly static spacetime with conformally flat and static slices}

Now we can discuss strictly static and vacuum spacetimes. 
The equation $R_{00}=-\alpha \tilde H_{00}$ gives us 
%
\begin{eqnarray}
D^2V=4 \frac{n-3}{n-1}\alpha D_i D_j V {}^{(n)}G^{ij}-\frac{\alpha}{n-1}V{}^{(n)}{\cal L}_{\rm GB} 
\end{eqnarray}
%
and then it is re-expressed as 
%
\begin{eqnarray}
D^i \Bigl(D_i V- 4 \frac{n-3}{n-1}\alpha D_j V {}^{(n)}G^{ij} \Bigr)
=-\frac{\alpha}{n-1}V{}^{(n)}{\cal L}_{\rm GB}. \label{div}
\end{eqnarray}
%
In the above we used the Bianchi identity. 

In the Einstein gravity, the above becomes $D^2V=0$. Then if the spacetime is strictly static, 
that is, there is the timelike and hypersurface orthogonal Killing vector everywhere, 
the volume integral of $D^2V=0$ directly implies that the ADM mass vanishes. Then the positive energy 
theorem implies that the spacetime is the Minkowski one. This statement holds in strictly stationary 
cases too. This is well-known fact \cite{Lich} (See Ref. \cite{SOS} for higher dimensional cases). 

Next we address the same issue in the Einstein-Gauss-Bonnet theory. In Eq. (\ref{div}), 
the right-hand side does not vanish. Here we consider some restricted cases, that is, 
$t=$constant hypersurfaces are conformally flat. And we employ the perturbative 
argument with respect to $\alpha$, that is, we will consider the field equations up to the 
linear order of $\alpha$. 
In Eq. (\ref{div}), this means that it is enough to estimate ${}^{(n)}{\cal L}_{\rm GB}$ in 
the zero-th order of $\alpha$. Since the slices are assumed to be conformally flat, the 
$n$-dimensional Weyl tensor vanishes, ${}^{(n)}C_{ijkl}=0$. 
The Hamiltonian constraint(Eq. (\ref{Hami})) gives us 
%
\begin{eqnarray}
{}^{(n)}R=O(\alpha).
\end{eqnarray}
%
The field equations $R_{ij}=-\alpha \tilde H_{ij}$ with Eq. (\ref{rij}) is approximated as 
%
\begin{eqnarray}
{}^{(n)}R_{ij}=\frac{1}{V}D_i D_j V+ O(\alpha).
\end{eqnarray}
%
Using of the above, we can compute ${}^{(n)}{\cal L}_{\rm GB}$ as 
%
\begin{eqnarray}
{}^{(n)}{\cal L}_{\rm GB}=-4\frac{n-3}{n-2}\frac{1}{V}D_i D_j V {}^{(n)}R^{ij}+O(\alpha).
\end{eqnarray}
%
Then Eq. (\ref{div}) becomes 
%
\begin{eqnarray}
D^i \Bigl( D_i V-4\frac{n-3}{n-2}\alpha {}^{(n)}R_{ij}D^j V \Bigr)=O(\alpha^2). 
\end{eqnarray}
%
Note that Eq. (\ref{div}) does not have the divergence-free form as the above if the $n$-dimensional 
Weyl tensor does not vanish. One may think that the imposing of the conformally flatness of $g_{ij}$ 
is too restrictive. But, it is still non-trivial to discuss the configuration of the 
strictly static vacuum spacetimes. 

The volume integral of the above becomes to the surface integral at the spatial infinity
%
\begin{eqnarray}
\int_{S_\infty} \Bigl( D_i V-4\frac{n-3}{n-2}\alpha {}^{(n)}R_{ij}D^j V \Bigr) dS^i=O(\alpha^2).\label{final}
\end{eqnarray}
%
This is because we assumed that there are no event horizons.  
In asymptotically flat spacetimes, 
near the spatial infinity($r=\infty$), ${}^{(n)}R_{ij}D^j V=O(1/r^{2n-1})$ and then 
we see that the second term in the integrand of Eq. (\ref{final}) does not contribute to 
the integral. Thus we can see that the ADM mass vanishes. 

The Ricci scalar of $g_{ij}$ is now computed as
%
\begin{eqnarray}
{}^{(n)}R=-\alpha {}^{(n)}{\cal L}_{\rm GB}=\frac{4(n-3)}{n-2}\alpha {}^{(n)}R_{ij}{}^{(n)}R^{ij}+O(\alpha^2). 
\end{eqnarray}
%
We know that the positive mass theorem holds if ${}^{(n)}R \geq 0$ is satisfied regardless of theories as 
long as we consider asymptotically flat spacetimes(the Riemannian positive mass theorem). 
We can see that ${}^{(n)}R$ is non-negative if $\alpha$ is positive. In the heterotic string theory, 
$\alpha$ is shown to be positive \cite{Metsaev}. Thus, the positive energy 
theorem holds. In the full order of $\alpha$, the positive mass theorem holds if 
$\alpha {}^{(n)}{\cal L}_{\rm GB} \leq 0$ is satisfied(See Ref. \cite{PMT-GB} for the related argument to the 
positive mass theorem in the Einstein-Gauss-Bonnet theory). But, we cannot expect that we can show that 
in general. After all, the positive mass theorem tells us that the static slice with the 
vanishing ADM mass is the Euclid space. Therefore, ${}^{(n)}R_{ij}=0$ holds and Eq. (\ref{div}) 
implies $D^2V=0$. These show us that $V$ is constant and the spacetime is indeed the Minkowski spacetime.

As a summary, we obtain the following statement:\\

 {\it Let us consider the asymptotically flat and strictly static spacetimes 
governed by the Einstein-Gauss-Bonnet theory with positive $\alpha$. 
We assume that the static slice is conformally flat. 
Up to the order $O(\alpha)$, then, it is turned out that the spacetime must be the Minkowski spacetime.}\\

There are several branches of the solutions in the Einstein-Gauss-Bonnet-theory. The condition of the 
asymptotic flatness may choose the Einstein-branch where the system reduces to the Einstein theory as 
$\alpha$ to zero. In such cases, it is natural to expect that the same features are shared with the 
Einstein theory.

\section{Summary and discussion}

In this paper we discussed the strictly static vacuum and asymptotically flat spacetimes in the 
Einstein-Gauss-Bonnet theory. We employed the perturbative argument with respect to the Gauss-Bonnet 
coupling constant $\alpha$. Here we considered the corrections up to the linear order of $\alpha$. 
Then we could show that the strictly static and conformally flat slices are the Euclid space for the cases with 
positive $\alpha$ and then the spacetime is the Minkowski spacetime. To show that, we used the positive 
mass theorem for the Einstein-Gauss-Bonnet theory. In the perturbative level, we saw that the Ricci 
scalar of the slices is non-negative and then the positive mass theorem indeed holds. 

There are a lot of remaining issues. In this paper we adopted the perturbative argument with respect to 
$\alpha$. Therefore one may consider the higher order terms of $\alpha$ or the Lovelock theory \cite{Lovelock}. 
We also focused on the strictly static vacuum spacetimes as a first step. It is nice to address 
the same issue in the strictly {\it stationary} spacetimes. In our discussion we used the 
Riemannian positive mass theorem which has been already proven. But, one might be able to 
improve/optimise the positive mass theorem for the Einstein-Gauss-Bonnet theory, that is, the condition of 
${}^{(n)}R \geq0$ may be relaxed. We are also interested in 
the black hole spacetimes. For example, we want to examine the uniqueness properties in the 
Einstein-Gauss-Bonnet theory. At least, one might be able to prove the black hole uniqueness at low energy 
scales.

\begin{acknowledgments}
We would like to thank Prof. Takashi Nakamura for his continuous encouragement. 
TS is supported by Grant-Aid for Scientific Research from Ministry of Education, Science,
Sports and Culture of Japan (No.~21244033).
SO is supported by JSPS Grant-in-Aid for Scientific Research (No. 23-855). 
The authors thank the Yukawa Institute for Theoretical Physics at Kyoto University, where 
this work was initiated during the YITP-T-12-04 on ``Nonlinear massive gravity theory and 
its observational test". 
\end{acknowledgments}



\end{document}